# Sweeping an oval to a vanishing point

Adrian Dumitrescu*     Minghui Jiang†

November 10, 2018

**Abstract**

Given a convex region in the plane, and a sweep-line as a tool, what is best way to reduce the region to a single point by a sequence of sweeps? The problem of sweeping points by orthogonal sweeps was first studied in [2]. Here we consider the following *slanted* variant of sweeping recently introduced in [1]: In a single sweep, the sweep-line is placed at a start position somewhere in the plane, then moved continuously according to a sweep vector $\vec{v}$ (not necessarily orthogonal to the sweep-line) to another parallel end position, and then lifted from the plane. The cost of a sequence of sweeps is the sum of the lengths of the sweep vectors. The (optimal) sweeping cost of a region is the infimum of the costs over all finite sweeping sequences for that region. An optimal sweeping sequence for a region is one with a minimum total cost, if it exists. Another parameter of interest is the number of sweeps.

We show that there exist convex regions for which the optimal sweeping cost cannot be attained by two sweeps. This disproves a conjecture of Bousany, Karker, O'Rourke, and Sparaco stating that two sweeps (with vectors along the two adjacent sides of a minimum-perimeter enclosing parallelogram) always suffice [1]. Moreover, we conjecture that for some convex regions, no finite sweeping sequence is optimal. On the other hand, we show that both the 2-sweep algorithm based on minimum-perimeter enclosing rectangle and the 2-sweep algorithm based on minimum-perimeter enclosing parallelogram achieve a $4/\pi \approx 1.27$ approximation in this sweeping model.

**Keywords**: Sweep-line, approximation algorithm, figure of constant width, convex polygon.

## 1 Introduction

The following question was raised by Paweł Żyliński [5]; see also [2]: Given a set of points in the plane, and a sweep-line as a tool, what is best way to move the points to a target point using a sequence of sweeps? The target point may be specified in advance or freely selected by the algorithm. In a single sweep, the sweep-line is placed in the plane at some start position, then moved orthogonally and continuously to another parallel end position, and then lifted from the plane. All points touched by the line are moved with the line in the direction of the sweep. When a point is swept over another point, and both are in the current set, the two points merge into one, and they are subsequently treated as one point. The cost of a sequence of sweeps is the total length of the sweeps, with no cost assessed for positioning or repositioning the line. Dumitrescu and Jiang have obtained several results on this question, among which we mention a ratio $4/\pi \approx 1.27$ approximation that uses at most 2 sweeps (or 4 sweeps if the target point is specified) and can be computed in $O(n \log n)$ time. We refer to this (original) model of sweeping as the *orthogonal sweeping model*.

---

*Department of Computer Science, University of Wisconsin–Milwaukee, WI 53201-0784, USA. Email: dumitres@uwm.edu. Supported in part by NSF CAREER grant CCF-0444188 and NSF grant DMS-1001667.

†Department of Computer Science, Utah State University, Logan, UT 84322-4205, USA. Email: mjiang@cc.usu.edu. Supported in part by NSF grant DBI-0743670.

Bousany et al. [1] have recently explored another variant, with points being replaced by a planar connected region, and orthogonal sweeps replaced by (possibly non-orthogonal) *slanted sweeps*. We refer to their model of sweeping as the *slanted sweeping model* or the *generalized sweeping model*: in a sweep operation the infinite sweep line may translate by any vector $\vec{v}$, not only by a vector orthogonal to the line; the corresponding cost is the (Euclidean) vector length $|\vec{v}|$. This model assumes the points swept have sufficient friction to avoid sliding along the sweep line. For convenience we also include segments in the class of regions; for instance the cost of sweeping a segment is its length. As for orthogonal sweeps, the sweep-line as a tool can be conveniently replaced by a finite sweep-segment of length twice the diameter of the point set.

Given a planar region, the *(optimal) sweeping cost* of the region is the infimum of the costs over all finite sweeping sequences of that region to a single point. An *optimal sweeping sequence* is one with a minimum total cost, if it exists. It is conceivable that the optimal sweeping cost of a region could be only approached in the limit, and not be attained through a finite sequence of sweeps. Another parameter of interest is the number of sweeps. We refer to a sequence of $k$ sweeps as a *$k$-sweep sequence*.

It is easy to exhibit non-convex planar regions whose optimal 2-sweep sequences are not optimal over all sequences [1]. Given a convex $n$-gon, Bousany et al. derived a linear-time algorithm for computing a minimum-perimeter parallelogram enclosing $P$, and thereby the optimal corresponding 2-sweep sequence [1]. They went further and conjectured that for planar convex regions, an optimal 2-sweep sequence makes in fact an optimal sweep sequence. Here we disprove this conjecture and thereby answer the main problem left open in [1].

**Theorem 1.** *There exist convex regions (such as the Reuleaux triangle, or a disk, or the isosceles trapezoid in Fig. 3) for which an optimal 2-sweep sequence is not optimal.*

We present two proofs of Theorem 1, based on different counterexamples. While the first proof is shorter, the second gives a better lower bound on the approximation ratio of the 2-sweep algorithm (minimum-perimeter enclosing parallelogram) from [1]. Moreover, the second proof also implies the existence of convex polygons with $n$ sides, for any $n \geq 4$, for which an optimal 2-sweep sequence is not optimal.

**Corrolary 1.** *For every $n \geq 4$ there exist convex polygons with $n$ vertices for which an optimal 2-sweep sequence is not optimal.*

In light of Theorem 1, one may naturally ask whether the 2-sweep algorithm of Bousany et al. has a good approximation ratio. In our previous paper based on the orthogonal sweeping model [2], we showed that a simple algorithm A2, which first computes a minimum-perimeter rectangle enclosing the given point set, then moves the point set to a single point by two orthogonal sweeps (four orthogonal sweeps if the target point is not freely chosen but is specified in the input), achieves an approximate ratio of $4/\pi$ in that model. Here we further show that the same algorithm A2 also achieves an approximation ratio of $4/\pi$ in the slanted sweeping model introduced in [1].

**Theorem 2.** *The 2-sweep Algorithm A2 (based on minimum-perimeter enclosing rectangle) gives a $4/\pi$-approximation for sweeping a (discrete or continuous) point set to a single point in the slanted sweeping model.*

Now for any point set $S$, the minimum-perimeter of a parallelogram enclosing $S$ is at most the minimum-perimeter of a rectangle enclosing $S$. Thus Theorem 2 implies that the approximation ratio of the 2-sweep algorithm of Bousany et al. is also at most $4/\pi$.

**Corrollary 2.** *The 2-sweep algorithm based on minimum-perimeter enclosing parallelogram [1] gives a $4/\pi$-approximation for sweeping a (discrete or continuous) point set to a single point in the slanted sweeping model.*

In the remainder of this paper, we restrict ourselves to the slanted sweeping model introduced in [1].

## 2 Proof of Theorem 1

We start with an upper bound on the cost of sweeping a convex polygon.

**Lemma 1.** *Let $P = A_1 \ldots A_n$ be a convex polygon of perimeter $\mathrm{per}(P)$, and let $s$ be an arbitrary side of $P$. Then the cost of sweeping $P$ is at most $\mathrm{per}(P) - |s|$.*

*Proof.* We can assume that $s = A_1 A_n$. Consider the triangulation of $P$ from the single point $A_n$; see Fig. 1. Make the first sweep with the line initially incident to $A_1$ and parallel to $A_2 A_n$ until the line is incident to $A_2 A_n$; the sweep vector is $\overrightarrow{A_1 A_2}$. Observe that after the sweep, we reduced the problem of sweeping $P = A_1 \ldots A_n$ to that of sweeping $P' = A_2 \ldots A_n$. We continue in a similar way: in the $i$th sweep ($i = 1, \ldots, n-2$), start with the line incident to $A_i$ and parallel to $A_{i+1} A_n$ until the line is incident to $A_{i+1} A_n$; the sweep vector is $\overrightarrow{A_i A_{i+1}}$. In the last sweep, $n-1$, we position the line incident to $A_{n-1}$ and (say) orthogonally to $A_{n-1} A_n$ and sweep until the line is incident to $A_n$. The polygon $P$ has been swept to the point $A_n$ in $n-1$ sweeps of total cost $|A_1 A_2| + \ldots |A_{n-1} A_n| = \mathrm{per}(P) - |A_1 A_n| = \mathrm{per}(P) - |s|$, as required. □

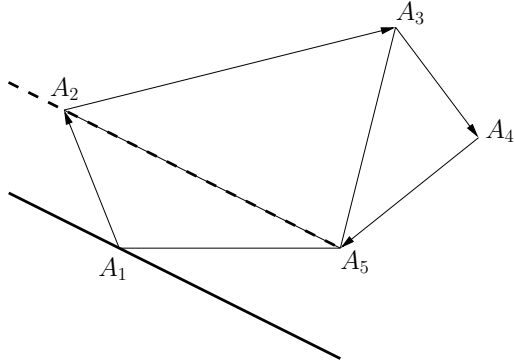

Figure 1: Sweeping a convex polygon; here $n = 5$.

**First proof of Theorem 1.** We first observe that the minimum-perimeter enclosing parallelogram of a *convex figure $\Theta$ of constant width $w$*, is a square whose side-length equals $w$. Indeed, assume the parallelogram $\Gamma$ encloses $\Theta$, a figure of constant unit width. Let $\ell_1, \ell'_1$ be a pair of parallel supporting lines of $\Gamma$, and let $\ell_2, \ell'_2$ be the other pair of parallel supporting lines of $\Gamma$. The distance between $\ell_1$ and $\ell'_1$ is 1, so the total length of the two parallel sides of $\Gamma$ along $\ell_2$ and $\ell'_2$ is at least 2, with equality if and only if $\ell_1 \perp \ell_2$. Similarly, the distance between $\ell_2$ and $\ell'_2$ is 1, so the total length of the two parallel sides of $\Gamma$ along $\ell_1$ and $\ell'_1$ is at least 2, with equality if and only if $\ell_1 \perp \ell_2$. Hence the unit square is the minimum-perimeter enclosing parallelogram of $\Theta$, and its semi-perimeter 2 is the conjectured sweep cost of $\Theta$ [1].

In the role of $\Theta$, consider a *Reuleaux triangle* $\Delta$ obtained from an equilateral unit triangle $ABC$ (a Reuleaux triangle can be obtained from an equilateral triangle $ABC$ by joining each pair of its vertices by a circular arc whose center is at the third vertex [4].) We can assume that $BC$ is horizontal. First sweep from $A$ orthogonally to $BC$ (the sweep-line is parallel to $BC$) until the sweep-line reaches $BC$. Observe that since the two upper tangents to $\Delta$ at $B$ and $C$ are vertical, the sweep reduces $\Delta$ to the circular cap $\Psi \subset \Delta$ based on $BC$. Let $B'C'$ be a slightly longer segment containing $BC$: $BC \subset B'C'$. Consider a slightly larger concentric circular cap based on $B'C'$ enclosing the cap $\Psi$ based on $BC$. Now make a fine equidistant subdivision of the circular arc connecting $B'$ and $C'$ to obtain a convex polygon $\Lambda$ enclosing the circular cap $\Psi$. By Lemma 1, the cost of sweeping $\Psi$ is at most the length of the arc $BC$ plus $\varepsilon$, for any given $\varepsilon > 0$. So the total cost of sweeping the Reuleaux triangle $\Delta$ is at most

$$\frac{\sqrt{3}}{2} + \frac{\pi}{3} + \varepsilon \leq 1.9133,$$

for a suitably small $\varepsilon$. Since this cost is smaller than 2, the 2-sweep conjecture is disproved. $\square$

**Remark.** A similar argument can be made for a disk $\Omega$ of unit diameter in the role of $\Theta$. Let $ABCD$ be an axis-parallel square enclosing $\Omega$, labeled counterclockwise and with $A$ as the lower left corner. As argued above, the minimum-perimeter enclosing parallelogram of $\Omega$ has semi-perimeter 2.

First sweep from $AB$ upward until the sweep-line is at distance $h$ (to be determined) from the top side $DC$ of the square. The sweep-line at this position cuts of a circular cap of $\Omega$ above the line, say between $D'$ and $C'$. Write $a = |D'C'|$. Make two sweeps with a vertical line, one from the left side $AD$ until it reaches $D'$, and one from the right side $BC$ until it reaches $C'$. Now sweep the circular cap based on $D'C'$ as in our proof for the Reuleaux triangle, and $\Omega$ has been swept to a single point. We parameterize $a = a(\alpha)$ and $h = h(\alpha)$ by the center angle $2\alpha$ of the circular cap. It is easy to see that $a(\alpha) = \sin\alpha$, $h(\alpha) = (1 - \cos\alpha)/2$, and the cost of our sweeping sequence is at most $2 - \big(a(\alpha) + h(\alpha) - \alpha\big) + \varepsilon$, for any given $\varepsilon > 0$. We are led to maximizing the function

$$f(\alpha) = a(\alpha) + h(\alpha) - \alpha = \sin\alpha + (1 - \cos\alpha)/2 - \alpha.$$

By simple calculus, it is easy to verify that $\alpha = 2\arcsin\frac{1}{\sqrt{5}}$ is the maximizing angle. Correspondingly, the cost of our sweeping sequence is at most

$$2 - \left(\frac{4}{5} + \frac{1}{5} - 2\arcsin\frac{1}{\sqrt{5}}\right) + \varepsilon = 1 + 2\arcsin\frac{1}{\sqrt{5}} + \varepsilon \leq 1.9273,$$

for a suitably small $\varepsilon$. Since this cost is smaller than 2, the same conclusion follows.

**Second proof of Theorem 1.** A useful geometric inequality that inspired our construction is the following lemma which is implicit in [1]; we have included here our own short proof for completeness (this fact was also known to us earlier).

**Lemma 2.** *Let $ABC$ be an obtuse triangle with $\angle BAC > \pi/2$, and $AD$ is the altitude. Put $a = |BC|$, $b = |AC|$, $c = |AB|$, and $h = |AD|$. Then $b + c < a + h$.*

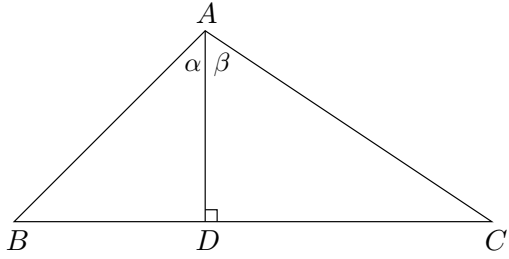

Figure 2: A lemma about obtuse triangles.

*Proof.* We refer to Fig. 2. Put $\alpha = \angle BAD$ and $\beta = \angle CAD$. Assume $0 < \alpha \leq \beta < \pi/2$ and $h = 1$. Then $a = |BD| + |CD| = \tan\alpha + \tan\beta$, $b = 1/\cos\beta$, $c = 1/\cos\alpha$, and $\pi/2 < \alpha + \beta < \pi$. We need to show that
$$1/\cos\alpha + 1/\cos\beta < 1 + \tan\alpha + \tan\beta. \tag{1}$$

We can prove (1) by tracing the following steps backwards:

$$\frac{1}{\cos\alpha} + \frac{1}{\cos\beta} < 1 + \frac{\sin\alpha}{\cos\alpha} + \frac{\sin\beta}{\cos\beta}$$
$$\cos\alpha + \cos\beta < \cos\alpha\cos\beta + \sin(\alpha+\beta)$$
$$1 - \sin(\alpha+\beta) < (\cos\alpha - 1)(\cos\beta - 1)$$
$$\left(\sin\frac{\alpha+\beta}{2} - \cos\frac{\alpha+\beta}{2}\right)^2 < \left(2\sin\frac{\alpha}{2}\sin\frac{\beta}{2}\right)^2$$
$$\sin\frac{\alpha+\beta}{2} - \cos\frac{\alpha+\beta}{2} < 2\sin\frac{\alpha}{2}\sin\frac{\beta}{2}$$
$$\sin\frac{\alpha+\beta}{2} < \cos\frac{\alpha+\beta}{2} + 2\sin\frac{\alpha}{2}\sin\frac{\beta}{2}$$
$$\cos\frac{\pi-\alpha-\beta}{2} < \cos\frac{\beta-\alpha}{2} \tag{2}$$
$$\pi - \alpha - \beta > \beta - \alpha \tag{3}$$
$$\pi/2 > \beta.$$

In the step from (3) to (2), the condition $\pi/2 < \alpha + \beta < \pi$ ensures that $0 < \pi - \alpha - \beta < \pi/2$, and the assumption $\alpha \leq \beta$ ensures that $\beta - \alpha \geq 0$. □

A polygon $Q$ is *flush* with an enclosed polygon $P$ if an edge $f$ of $Q$ contains an edge $e$ of $P$, i.e., $e \subseteq f$; the two edges $e$ and $f$ are then said to be flush with each other.

Another useful tool is the following lemma of Bousany et al. [1, Lemma 4] that was implied by the proof of a lemma of Mitchell and Polishchuk [3, Lemma 1]:

**Lemma 3.** *A minimum-perimeter parallelogram $Q$ enclosing a convex polygon $P$ has one edge flush with an edge of $P$.*

We next present another counterexample (a trapezoid) to the 2-sweep conjecture of Bousany et al. [1]. We start with an intuitive overview. Refer back to Fig. 2 for Lemma 2. Suppose that the triangle $ABC$ is isosceles, i.e., $|AB| = |AC|$. Then by Lemma 3 we can show that this triangle has only two candidates for a minimum-perimeter parallelogram. The first candidate is a rectangle flush with $BC$; the second candidate is a parallelogram (indeed a rhombus) flush with both $AB$ and $AC$. By Lemma 2, the first candidate has a larger perimeter than the second.

Refer now to Fig. 3. We shave off the vertex $A$ of the triangle $ABC$ by an edge $EF$ parallel to $BC$, and obtain a trapezoid $EBCF$. By setting the angles of the triangle and the amount of cutting to some suitable values, we can show that the trapezoid obtained from the triangle still have two candidates for a minimum-perimeter parallelogram: a smaller rectangle flush with $BC$ and $EF$, and the same parallelogram flush with both $AB$ and $AC$. With the right amount of cutting, the perimeters of the two candidates can be made the same. Thus the optimal 2-sweep sequence for this trapezoid has a cost of $|AB| + |AC|$. On the other hand, the method in Lemma 1 gives a 3-sweep sequence of cost $|BE|+|EF|+|FC|$. By triangle inequality, we have $|EF| < |AE|+|AF|$, and hence $|BE| + |EF| + |FC| < |BE| + |AE| + |AF| + |FC| = |AB| + |AC|$.

PSfrag replacements

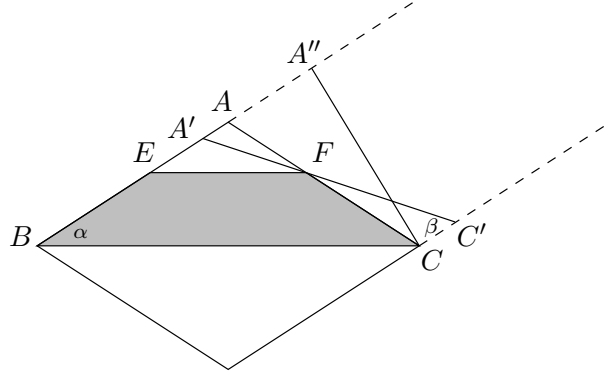

Figure 3: A trapezoid $EBCF$ cut from an isosceles triangle $ABC$. $|BC| = 2$, $\angle ABC = \angle ACB = \alpha$. $|EF|/|BC| = \kappa = \cos 2\alpha$. $\angle A'C'C = \beta$.

We now give a more precise description of the construction. Refer to Fig. 3. We start with an isosceles triangle $ABC$ with $|BC| = 2$ and $\angle ABC = \angle ACB = \alpha$, where $\alpha = \arcsin x$ and $x = ((\sqrt{297} + 17)^{1/3} - (\sqrt{297} - 17)^{1/3} - 1)/3 = 0.5436\ldots$ is the unique real root of the cubic equation $x^3 + x^2 + x - 1 = 0$. A calculation shows that $\alpha = 32.9351\ldots°$. The parameter $\alpha$ is chosen in this way to satisfy the following equation:

$$2 + (1 - \cos 2\alpha) \tan \alpha = 2/\cos \alpha. \tag{4}$$

Put $\kappa = \cos 2\alpha$. A calculation shows that $\kappa = 0.4088\ldots$. We cut a smaller isosceles triangle $AEF$ off the triangle $ABC$, where $EF \parallel BC$ and $|EF|/|BC| = \kappa$. Thus we obtain a trapezoid $EBCF$. We next consider the minimum-perimeter enclosing parallelograms of this trapezoid.

Let $\Gamma$ be a minimum-perimeter parallelogram (which may not be unique) of the trapezoid $EBCF$. By Lemma 3, $\Gamma$ has one edge flush with one of the four edges of $EBCF$. There are two cases.

*Case 1*: $\Gamma$ has an edge flush with either $BC$ or $EF$. Then $\Gamma$ must be the enclosing rectangle of $EBCF$ with $BC$ as one edge and with the opposite parallel edge flush with $EF$. The altitude of $ABC$ from $A$ to $BC$ is $h = (|BC|/2) \tan \alpha = \tan \alpha$. By the similarity of the triangles $ABC$ and $AEF$, the distance between the two parallel lines along $BC$ and $EF$ is $(1-\kappa)h = (1-\cos 2\alpha) \tan \alpha$. The semi-perimeter of $\Gamma$ is $|BC|+(1-\kappa)h = 2+(1-\cos 2\alpha) \tan \alpha$, the left-hand side of Equation (4).

*Case 2*: $\Gamma$ has an edge flush with either $BE$ or $CF$. By symmetry, we can assume that $\Gamma$ has an edge flush with $BE$. Then the opposite parallel edge must go through the point $C$. Now consider the other two parallel edges of $\Gamma$. By the minimality of $\Gamma$, one of these two edges, say $e$, must go through the point $C$, $F$, or $E$. If the edge $e$ goes through the point $C$, then we can rotate $e$ counter-clockwise about $C$, from $CA''$ to $CA$, until it also goes through $F$. Note that $BA'' + CA'' = BA + AA'' + CA'' \geq BA + CA$. The semi-perimeter of $\Gamma$ can only decrease during this rotation. Analogously, if $e$ goes through $E$, then we can rotate $e$ clockwise about $E$ until it also goes through $F$, without increasing the semi-perimeter of $\Gamma$. Therefore, we can assume that $\Gamma$ has an edge $e = A'C'$ going through the point $F$. Let $\beta$ be the angle of $\Gamma$ between the edge $A'C'$ and the adjacent edge through $C$. Then $\alpha \leq \beta \leq 2\alpha$: $A'C'$ is flush with $EF$ when $\beta = \alpha$, and is flush with $CF$ when $\beta = 2\alpha$. We next show that the semi-perimeter of $\Gamma$, denoted $\ell(\beta)$, is minimized when $\beta = 2\alpha$.

By the similarity of the triangles $AFA'$ and $CFC'$, we have $|A'F|/|C'F| = |AF|/|CF|$ and hence $|A'F|/|A'C'| = |AF|/|AC|$. By the similarity of the triangles $AEF$ and $ABC$, we have $|AF|/|AC| = |EF|/|BC|$. Thus $|A'F|/|A'C'| = |EF|/|BC| = \kappa = \cos 2\alpha$. Since $|BA| + |AF|\cos 2\alpha = |BA'| + |A'F|\cos\beta$, we have

$$|BA'| = |BA| + |AF|\cos 2\alpha - |A'F|\cos\beta$$
$$= |BA| + |AF|\cos 2\alpha - |A'C'|\cos 2\alpha \cos\beta.$$

Also, since $|AC|\sin 2\alpha = |A'C'|\sin\beta$, we have

$$|A'C'| = |AC|\sin 2\alpha / \sin\beta.$$

It follows that

$$\ell(\beta) = |BA'| + |A'C'| = |BA| + |AF|\cos 2\alpha + |A'C'|(1 - \cos 2\alpha \cos\beta)$$
$$= |BA| + |AF|\cos 2\alpha + |AC|\sin 2\alpha \frac{1 - \cos 2\alpha \cos\beta}{\sin\beta},$$

and that

$$\frac{\mathrm{d}}{\mathrm{d}\beta}\ell(\beta) = |AC|\sin 2\alpha \frac{\cos 2\alpha \sin\beta \cdot \sin\beta - (1 - \cos 2\alpha \cos\beta)\cos\beta}{\sin^2\beta}$$
$$= |AC|\sin 2\alpha \frac{\cos 2\alpha - \cos\beta}{\sin^2\beta}.$$

Since $\alpha \leq \beta \leq 2\alpha < \pi/2$, we have $\frac{\mathrm{d}}{\mathrm{d}\beta}\ell(\beta) \leq 0$. Thus $\ell(\beta)$ is minimized when $\beta = 2\alpha$. This implies that $\Gamma$ is a rhombus with both $AB$ and $AC$ as edges. Since $|AB| = |AC| = (|BC|/2)/\cos\alpha = 1/\cos\alpha$, the semi-perimeter of $\Gamma$ is $|AB| + |AC| = 2/\cos\alpha$, the right-hand side of Equation (4).

In either Case 1 or Case 2, by Equation (4), the semi-perimeter of a minimum-perimeter enclosing parallelogram of the trapezoid $EBCF$ is $2 + (1 - \cos 2\alpha)\tan\alpha = 2/\cos\alpha = 2.3829\ldots$, which is the minimum cost of any 2-sweep sequence. On the other hand, by Lemma 1, there is a 3-sweep sequence of cost $|BE| + |EF| + |FC|$. Recall that $|BC| = 2$ and $|AB| = |AC| = 1/\cos\alpha$. Since $|EF|/|BC| = \kappa = \cos 2\alpha$, we have $|EF| = 2\kappa = 2\cos 2\alpha$ and $|BE| = |FC| = (1 - \kappa)/\cos\alpha = (1 - \cos 2\alpha)/\cos\alpha$. Then $|BE| + |EF| + |FC| = 2\cos 2\alpha + 2(1 - \cos 2\alpha)/\cos\alpha = 2.2264\ldots$. The ratio of the two costs is $1.0703\ldots$. This concludes the second proof of Theorem 1. □

**Remark.** In the construction we just described, the parameter $\kappa$ is set to $\cos 2\alpha$ so that the semi-perimeter of the parallelogram in Case 2 is a decreasing function of $\beta$, which ensures that the parallelogram is in fact a rhombus flush with both $AB$ and $AC$. The parameter $\alpha$ is then determined by balancing the two cases as in Equation (4). If we relax the rhombus requirement, then the analysis will become more complex. With the help of a computer program, we verified that when $\alpha = 35.478\ldots°$ and $\kappa = 0.3614\ldots$ (here $\kappa > \cos 2\alpha$), the ratio of the minimum cost of any 2-sweep sequence to the cost of the 3-sweep sequence for the resulting trapezoid is $1.0715\ldots$. This gives a lower bound of $1.0715\ldots$ on the approximation ratio of the algorithm from [1].

It is likely that the lower bound may be improved further. Consider a circular cap $\Psi$ of center angle $2\alpha$, where $\alpha = \arctan(1/2)$. Let $\Psi$ be subtended by a chord $AB$ in a unit diameter disk. Observe that $\alpha$ satisfies the equation

$$\sin\alpha + (1 - \cos\alpha)/2 = \tan\alpha.$$

We suspect that (for this $\alpha$) there are exactly two minimum-perimeter enclosing parallelograms for $\Psi$, namely, a rectangle based on $AB$, and a rhombus with diagonal $AB$. Then an easy calculation shows that the ratio between the semi-perimeter of these parallelograms and the upper bound $\alpha+\varepsilon$ on the sweeping cost is at least $\frac{\tan\alpha}{\alpha+\varepsilon}$, for any $\varepsilon > 0$; or equivalently, at least $\frac{1}{2\arctan(1/2)} - \varepsilon = 1.0784\ldots - \varepsilon$, for any $\varepsilon > 0$. When $\varepsilon$ tends to zero this ratio is at least $\frac{1}{2\arctan(1/2)} = 1.0784\ldots$.

## 3 Proof of Theorem 2

We refer to [2] for a description of our Algorithm A2 and its analysis in the orthogonal sweeping model. The analysis bounds the effect of each sweep on the current point set, in terms of the reduction in semi-perimeter of the minimum enclosing rectangle in every orientation $\beta$. We then integrate the total effect of all sweeps in an optimal (or nearly optimal solution) over all orientations. That is, we consider an arbitrary $k$-sweep sequence $\sigma$ given by the vectors $\vec{v}_i$, $i = 1,\ldots,k$, where $x_i = |\vec{v}_i|$ and $\alpha_i$ is the length and respectively, the direction (angle) of $\vec{v}_i$. Let $\gamma_i$, $i = 1,\ldots,k$, be the direction (angle) of the sweep-line in the $i$th sweep. The cost of $\sigma$ is $x = \sum_{i=1}^{k} x_i$. In the end we let $x \leq \mathrm{OPT} + \varepsilon$, where OPT is the optimal sweeping cost (in the analysis in [2] we overlooked the possibility that no finite sweeping sequence is optimal; this only introduces the correction by $\varepsilon$ in the calculation, see below).

Fix an orientation $\beta$, and consider the minimum enclosing rectangle $Q_i(\beta)$ of the current point-set in this orientation just before the $i$th sweep. To adapt the analysis from the orthogonal sweeping model to the slanted sweeping model, we only need to show, as in [2], that the reduction in the semi-perimeter of $Q_i(\beta)$ (to that of $Q_{i+1}(\beta)$ in the next step) due to the $i$th slanted sweep is still bounded from above by $x_i(|\cos(\alpha_i - \beta)| + |\sin(\alpha_i - \beta)|)$, the expression on the left-hand side of equation (1) in [2].

Observe that in the $i$th sweep, any point swept moves in the direction $\alpha_i$ by at most $x_i$, regardless of the sweep-line orientation $\gamma_i$. Thus the projections of this move onto the two orthogonal directions $\beta$ and $\beta + \pi/2$ are at most $x_i|\cos(\alpha_i - \beta)|$ and $x_i|\sin(\alpha_i - \beta)|$, respectively. Hence their sum is at most $x_i(|\cos(\alpha_i - \beta)| + |\sin(\alpha_i - \beta)|)$, and consequently, the reduction in the semi-perimeter of $Q_i(\beta)$ is bounded by the same quantity, as claimed. Analogous to [2], by integration we find that the approximation ratio of the 2-sweep algorithm is bounded from above by $4/\pi + \varepsilon$, for any $\varepsilon > 0$. By letting $\varepsilon$ tend to zero, we conclude that this ratio is at most $4/\pi$, as required.

## 4 Conclusion

We have shown the existence of convex regions whose optimal 2-sweep sequences are not optimal overall. Concerning the number of sweeps of optimal sweeping sequences, we don't think that the number 2 (in Conjecture 3 from [1]) has any special role. As a tentative example consider a circular cap of small center angle, say $\alpha \leq \pi/18$; see Fig. 4. In fact our counterexamples in Section 2 were all suggested by the circular cap.

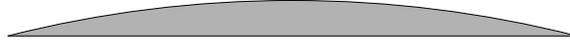

Figure 4: A circular cap of small angle.

**Conjecture 1.** *There exist convex regions (such as the circular cap in Fig. 4) for which no finite sweeping sequence is optimal.*

Near the end of their paper, Bousany et al. [1] suggested that one small step toward proving their 2-sweep conjecture would be to prove that three sweeps are never needed for triangles. Recall that the convex region in our second proof of Theorem 1 is a trapezoid, which shows that their conjecture is already false for quadrilaterals. Moreover, for every $n \geq 5$ it is not hard to modify the trapezoid example (by making a slight perturbation and by adding extra vertices) into a convex polygon with $n$ vertices for which no 2-sweep sequence is optimal (thereby we obtain Corollary 1). Whether the 2-sweep conjecture holds for triangles remains open.